\documentstyle[aps,prl,multicol,graphicx]{revtex}

\begin{document}

\author{A. Morozov, B. Horovitz}

\address{Department of Physics and Ilze Katz center for
nanotechnology, Ben Gurion university, Beer Sheva 84105 Israel}

\author{P. Le Doussal}
 \address{CNRS-Laboratoire de Physique Th{\'e}orique
de l'Ecole Normale Sup{\'e}rieure, 24 rue Lhomond,75231 Cedex 05,
Paris France.}

\title{Decoupling and Decommensuration in Layered Superconductors
with Columnar Defects}
\maketitle

\begin{abstract}
We consider layered superconductors with a flux lattice
perpendicular to the layers and random columnar defects parallel
to the magnetic field B. We show that the decoupling transition
temperature $T_d$, at which the Josephson coupling vanishes, is
enhanced by columnar defects by an amount $\delta T_d/T_d\sim
B^2$. Decoupling by increasing field can be followed by a
reentrant recoupling transition for strong disorder. We also
consider a commensurate component of the columnar density and show
that its pinning potential is renormalized to zero above a
critical long wavelength disorder. This decommnesuration
transition may account for a recently observed kink in the melting
line.
\end{abstract}


\begin{multicols}{2}

\vspace{1cm}
The phase diagram of layered superconductors in a magnetic field
$B$ perpendicular to the layers is of considerable interest in
view of recent experiments on high temperature superconductors
\cite{kes,fuchs,marley,bruyn}. Columnar defects (CD) induced by
heavy ion bombardment provide an additional interesting probe
\cite{kes}. In particular the irreversibility line at low
temperatures is enhanced \cite{klein,seow} while within the liquid
phase an onset of enhanced $z$ axis correlation
\cite{kosugi,sato,pomar} was observed. Recent data \cite{zeldov}
indicates that CD produce a porous vortex matter in which ordered
vortex crystallites are embedded in the 'pores' of a rigid matrix
of vortices pinned on the CD's. A sharp kink in the melting curve
signals an abrupt change from melting enhanced by the matrix at
high fields to a more weakly enhanced melting at lower fields.
Theoretical studies on CD's have shown a "localization" transition within a
bose glass phase \cite{nelson,koshelev2}.
Recent simulations \cite{hu} have been interpreted in terms of a
Bragg-Bose glass with positional order which sets in as field
increases. Also a "recoupling" crossover transition
\cite{bulaevskii} was studied in the vortex liquid phase.

In the absence of CD theoretical studies have shown layer
decoupling due to thermal fluctuations
\cite{glazman,daemen,horovitz1} or due to disorder
\cite{horovitz1,koshelev}. At this phase transition the Josephson
coupling between layers vanishes at long scales, i.e. the critical
current perpendicular to the layers vanishes and superconducting
correlations in the $z$ direction (perpendicular to the layers)
become short range. Decoupling involves in principle also
proliferation of point defects - vacancies and interstitials (VI)
\cite{HL}. The flux lattice is present even in the decoupled phase
with the $z$ axis positional correlations maintained by magnetic
couplings. In the case of point disorder this phase would thus
still exhibit Bragg glass type order without dislocations.

An increase in the critical current at a "second peak" transition
has been interpreted as due to an apparent discontinuity in the
tilt modulus at decoupling \cite{horovitz2}. Plasma resonance data
\cite{shibauchi,matsuda} has shown a significant jump at this
transition, consistent with the decoupling scenario. Whether this
transition is driven by decoupling alone rather than by a sudden
dislocation proliferation \cite{GL} remains to be investigated.

In the present work we consider the effects of CD within the flux
lattice phase, neglecting VI (whose role is discussed below). We
find that the decoupling transition temperature $T_d(B)$ is
enhanced by CD. In particular for strong disorder the low field
form $T_d(B)\sim 1/B$ becomes $T_d(B)\sim B$ at strong fields,
hence decoupling followed by a reentrant transition into a coupled
state, i.e. recoupling, is possible with increasing field. These
predictions can test wether the "second peak" transition is of a
decoupling type. We also allow for a finite component of the CD
density which is commensurate with the flux lattice, a component
which usually needs to be specifically prepared
\cite{moshchalkov}. We find that long wavelength disorder
renormalizes the commensurate coupling to zero, i.e.
decommensuration, above a critical value of disorder. We propose
that the matrix component of the porous vortex matter provides a
commensurability potential for the embedded crystallites. At high
fields this enhances the crystallite melting temperature, while below the
decommensuration transition the crystallites decouple from the
matrix, leading to a weaker enhancement of melting.

We study the classical partition function of $L/d$ Josephson
coupled layers where $L\rightarrow \infty$ is the total length in
the $z$ direction perpendicular to the layers and $d$ is the
interlayer spacing. The elastic energy of the transverse
displacement fields $u(q,k)$ in the absence of Josephson coupling
can be written as \cite{sudbo,goldin}
\begin{equation}\label{el}
H_{el} = \frac{1}{2L} \sum_{k,a} \int_q (c_{66} q^2 + c_{44}^0(k)
k_z^2) |u^a(q,k)|^2
\end{equation}
where a replica index $a=1,2...,n$ is needed below for the
disorder average. The elastic constants are \cite{sudbo,goldin}
$c_{44}^0(k) = \tau/(8 d a_0^2 \lambda_{ab} ^2 k_z^2) \ln(1 +
a_0^2 k_z^2/4 \pi)$, $c_{66} = \tau/(16 d a_0^2)$, where $k$ is
the wavevector in the $z$ direction, $k_z=(2/d) \sin(k d/2)$,
$\tau=\Phi_0^2 d/(4\pi^2\lambda_{ab}^2)$, $\lambda_{ab}$ is the
magnetic penetration length parallel to the layers, $a_0^2$ is the
area of the flux lattice unit cell and $\Phi_0$ is the flux
quantum, i.e. $\Phi_0=Ba_0^2$ . Note that the Josephson coupling
induces an additional term in $c_{44}$ \cite{goldin}, as also
shown below. The decoupling transition of the pure system (for
weak Josephson coupling) is given by \cite{daemen,horovitz1}:
\begin{equation}\label{Td0}
T_d^0 = \frac{4 a_0^4}{d^2} (\int_k \frac{dk}{c_{44}^0(k)})^{-1}
\end{equation}
and our principal aim is to obtain the corresponding $T_d$ in
presence of correlated disorder.

Consider a distribution of CD whose positions within a layer are
random and uncorrelated. Each of the CD has a radius $b_0$ and
their average areal density $n_{CD}$ is low, $n_{CD}b_0^2\ll 1$. A
flux line has a core of radius $\xi_0$ which usually satisfies
$\xi_0<b_0$. Once a flux line is partially inside a CD it gains
its core energy $E_c$ per layer. The pinning potential per unit
area is then $U_{pin}({\bf r})=(E_c/\xi_0^2)\sum_i p({\bf r}-{\bf
r}_i)$ with the sum on the CD positions and $p({\bf r})$ is a
shape function, e.g. $p({\bf r})=1$ for $r<b_0$ and vanishes for
$r>b_0$. The variance, neglecting CD overlaps, is therefore
\begin{equation}\label{Upin}
\overline{U_{pin}({\bf r})U_{pin}({\bf r}')}\approx
E_c^2n_{CD}(b_0/\xi_0)^4\delta^2({\bf r}-{\bf r}') \,.
\end{equation}
The average with respect to a flux density involves an additional
factor $(\xi_0/a_0)^4$ due to the decomposition of a sharply
peaked flux into harmonics with reciprocal vectors ${\bf Q}$. The
replica average at temperature $T$ is then \cite{giam}
\begin{eqnarray}\label{dis}
{\cal H}_{dis} && =- \sum_{ab} \{ \frac{1}{L} \sum_k \int
\frac{d^2q}{(2\pi)^2} s q^2 L \delta_{k,0} u^a({\bf q}
,k) u^b(-{\bf q},-k) \nonumber\\
&&+ W \sum_{n,n'} \int d^2r \cos[{\bf Q} \cdot ({\bf u}^a_n({\bf
r} ) - {\bf u}^b_{n'}({\bf r}))] \}/2T
\end{eqnarray}
where $W=E_c^2n_{CD}(b_0/a_0)^4$ and only the shortest most
relevant \cite{giam} ${\bf Q}$ is retained. The $\cos$ term above
involves vectors ${\bf Q}$ and ${\bf u}_n^a$ which in the averages
below yield $\langle {\bf Q}\cdot {\bf u}\rangle ^2= Q^2\langle
u^2+u_{l}^2\rangle/2$; $u_{l}$ is the longitudinal displacement
which is reconsidered below, but is neglected for now as it has no
effect on the decoupling transition. The parameter $s$ measures a
long wavelength random torque coupled to a local bond angle
\cite{carpentier} $\gamma=(\partial_xu_y-\partial_yu_x)/2$  since
for transverse modes $({\mbox{\boldmath $\nabla$}}u)^2=4\gamma^2$;
note that the usual long wavelength disorder couples to
${\mbox{\boldmath $\nabla$}}\cdot {\bf u}$, hence it involves only
longitudinal modes.

The long range Bragg glass properties depend on the nonlinear
$\cos$ term in Eq. (\ref{dis}). If this $\cos$ is expanded, it
yields $\sim \sum_{ab}\int d^2r u^a({\bf r},k=0)u^b({\bf r},k=0)$,
i.e. a $k=0$ quadratic term which has no effect on the decoupling
transition. It is therefore essential to treat the Bragg glass
nonlinearities properly.

We also allow for a commensurate term of the CD density of the
form
\begin{equation}\label{com}
{\cal H}_{com}=-y_c(2d/Q^2)\sum_{n,a}\int d^2r \cos [{\bf
Q}\cdot {\bf u} _n^a({\bf r})]\,.
\end{equation}
This term assumes a predesigned component of the CD density, 
in addition to the random one.

Consider next the Josephson phase, i.e. the relative
superconducting phase of two neighboring layers. Each flux line
can be viewed as a collection of point singularities, or pancake
vortices, positioned one on top of the other in consecutive
layers. Around each pancake vortex the superconducting phase
follows the angle $\alpha({\bf r})$ which changes by $2\pi$ in a
complete rotation. The Josephson phase involves then a nonsingular
component $\theta _n({\bf r})$ and a singular contribution from
pancake vortices. The latter are positioned at ${\bf R}_l+{\bf
u}_l^n$ in the $n$-th layer and at ${\bf R}_l+{\bf u}_l^{n+1}$ in
the $(n+1)$-th layer, where ${\bf R}_l$ is the undistorted
position of the $l$-th flux line. The total Josephson phase is
then
\begin{eqnarray}
&&\theta _n({\bf r})+\sum_l[\alpha({\bf r}-{\bf R}_l-{\bf u}_l^n)-
\alpha({\bf r}-{\bf R}_l-{\bf u}_l^{n+1})]\nonumber\\
&&\approx \theta _n({\bf r})+\sum_l({\bf u}_l^n-{\bf
u}_l^{n+1}){\mbox{\boldmath $\nabla$}} \alpha({\bf r}-{\bf R}_l)
\end{eqnarray}
where the expansion is justified in the Bragg glass since the
correlation length in the $z$ direction is $\gg d$. We define
(including now the replica index) $b^a({\bf q},k) = -2 \pi d e^{i
k d/2} u^a({\bf q},k)k_z/(qa_0^2)$
so that the Josephson phase is $\theta^a_n({\bf r}) + b^a_n({\bf
r})$. Fluctuations of the $\theta_n({\bf r})$ field involve the
Josephson energy as well as magnetic field terms,
\begin{eqnarray}\label{Jos}
{\cal H}_{J} &&= \frac{1}{2 L} \sum_{k,a} \int
\frac{d^2q}{(2\pi)^2} G_f^{-1}(q,k) |\theta^a ({\bf q},k)|^2
\nonumber\\
&&- y_J \sum_{n,a} \int d^2r \cos[\theta^a_n({\bf r}) + b^a_n({\bf
r})]
\end{eqnarray}
where \cite{horovitz3} $G_f(q,k)= 4 \pi d^3 (\lambda_{ab}^{-2} +
k_z^2)/(\tau q^2)$. The full Hamiltonian is then ${\cal H}={\cal
H}_{el}+{\cal H}_{dis}+{\cal H}_{com}+{\cal H}_J $.

We proceed to solve this system by the variational method allowing
for replica symmetry breaking (RSB) \cite{giam,mezard}. The form
of the variational Hamiltonian ${\cal H}_0$ is obtained by
expanding the $\cos$ terms and replacing $y_J$, $y_c$ and $W$ by
variational parameters $z_J$, $z_c$ and $\sigma_{ab}(k)$,
respectively. The Josephson term involves a $\theta\,, b$ cross
term which is eliminated by a shift $\tilde{\theta}^a({\bf
q},k)=\theta^a({\bf q},k)-u^a({\bf q},k)z_J(2\pi
k_z/a_0^2q)\exp(ikd/2)/(G_f^{-1}+z_J/d)$. Hence (repeated indices
are summed)
\begin{eqnarray}\label{H0}
&{\cal H}_0=& \frac{1}{2L} \sum_k \int \frac{d^2q}{(2\pi)^2}
\{G_{ab}^{-1}(q,k)u^a({\bf q},k) u^{b*}({\bf q},k)\nonumber\\
&&+[G_f^{-1}(q,k)+z_J/d]|\tilde{\theta}^a({\bf q},k)|^2 \}\,,\\
 &G_{ab}^{-1}(q,k)=&[c_{66}q^2+c_{44}(q,k)k_z^2
+z_c]\delta_{ab}\nonumber\\
&&-sL\frac{q^2}{T}\delta_{k,0}-\sigma_{ab}(k)\label{G}\,.
\end{eqnarray}
The effect of the nonsingular $\theta^a$ is to shift $c_{44}^0(k)$
of Eq. (\ref{el}) into $c_{44}(q,k)$,
\begin{equation}\label{c44}
c_{44}(q,k)=c_{44}^0(k)+\frac{B^2}{4\pi
(1+\lambda_c^2q^2+\lambda_{ab}^2k_z^2)}
\end{equation}
where $\lambda_c^2=\Phi_0^2/(16\pi^3z_Jd)$. Note that the limit
$\lambda_c\rightarrow \infty$ at decoupling must be taken before
$q\rightarrow 0$.

 The variational method minimizes the free energy $F_0+\langle
{\cal H}-{\cal H}_0\rangle_0$ where the free energy $F_0$ and the
average $\langle ...\rangle_0$ correspond to ${\cal H}_0$. This
yields
\end{multicols}
\begin{eqnarray}
\sigma_{ab}(k)&=& (WQ^2/2d^2T)\int_0^L dz[\cos kz\exp
(-B_{ab}(z)/2)-\delta_{ab}\sum_c\exp(-B_{ac}(z)/2)]\label{self1}\\
 z_J&=& y_J \exp \{- (T/2) \int_{q,k} [
\frac{k_z^2}{q^2}(\frac{2\pi d}{a^2})^2
(\frac{G_f^{-1}(q,k)}{G_f^{-1}(q,k)+z_J/d})^2G_{aa}(q,k) +
(G_f^{-1}(q,k)+z_J/d)^{-1} ]\}\label{self2}\\
z_c &=& y_c\exp \{-(TQ^2/4) \int_{q,k}G_{aa}(q,k)\}\label{self3}
\end{eqnarray}
\begin{multicols}{2}
where $\int_{q,k}=\int d^2qdk/(2\pi)^3$ and  $B_{ab}(z)$ is given
by
\begin{eqnarray}\label{B}
B_{ab}(z)=T Q^2 \int \frac{d^2qdk}{(2\pi)^3}[G_{aa}(q,k)-\cos (kz)
G_{ab}(q,k)]\nonumber\,.
\end{eqnarray}
Since the disorder is $z$ independent the off diagonal terms
$B_{a\neq b}(z)$ are $z$ independent so that $\sigma_{a\neq b}$
has only a $k=0$ component; hence RSB is present only at $k=0$. It
is convenient to define $G_c^{-1}(q,k)=\sum_bG_{ab}^{-1}(q,k)$ so
that for $k\neq 0$ $G_c(q,k)=G_{aa}(q,k)$. The RSB solution
reduces here to a one step form \cite{giam}, hence $G_c(q,k)$ can
be written with self energies in the form
\begin{equation}\label{Gc}
G_c^{-1}(q,k)=c_{66}q^2+c_{44}(q,k)k_z^2 +z_c
+\Sigma_1(1-\delta_{k,0}) +I(k)
\end{equation}
\begin{equation}\label{Sigma1}
\Sigma_1+z_c=(WQ^4/16\pi d^2 c_{66})\exp (-B_+/2)\,.
\end{equation}
$B_+=T Q^2 \int_{q,k}G_c(q,k)$  is a Debye Waller factor which is
dominated by large $q,k$ so that $c_{44}^0(k)$ can be used to
obtain
\begin{equation}\label{B+}
|B_+|\lesssim \frac{16\pi T}{\tau}\ln
\frac{c_{66}Q^2}{(\tau/8da_0^2\lambda_{ab}^2)+\Sigma_1+I(\pi/d)+z_c}
\end{equation}
while the function $I(k)$ satisfies for $T\ll \tau$
\begin{eqnarray}\label{I}
I(k)=&&4\pi c_{66}(\Sigma_1+z_c)\int \frac
{d^2q}{(2\pi)^2}[(c_{66}q^2+z_c+\Sigma_1)^{-1}\nonumber\\
&&-G_c(q,k)] \,.
\end{eqnarray}
Note that for $k\rightarrow 0$ this yields $I(k)\sim |k|$ while
 $I(k)\approx \Sigma_1+z_c$ for large $k$, up to logarithmic terms.
A condition for melting can be estimated by a Lindemann number
$c_L\approx 0.15$ \cite{nelson} so that $\langle u_n^2({\bf r})
/a_0^2\rangle=B_+/4\pi ^2\approx c_L^2$. Hence $\tau$ is a measure
of the melting temperature and for $T\ll \tau$ $B_+$ is small. We
note that it is essential to keep the nonsingular phase $\theta$
to obtain the correct structure factor $G_c(q,k)$ in Eq.
(\ref{Gc}).

The decoupling transition is determined by the vanishing of $z_J$.
Eq. (\ref{self2}) can be written in the form
\begin{eqnarray}\label{z}
z_J=y_Je^{-\frac{T}{2}\int_{q,k}[\frac{z_J}{d}+ [G_f(q,k)+
(\frac{2\pi dk_z}{a_0^2q})^2
G_c(q,k;z_J=0)]^{-1}]^{-1}}\nonumber\,.
\end{eqnarray}
 The integral is dominated by
$k\gg q$ so that for $k>1/\lambda_{ab}$ the effect of the $\theta$
field via $G_f(q,k)$ is negligible for $a_0^2/8\pi
\lambda_{ab}^2\ll 1$. The $q$ integration is then dominated by
$c_{44}^0(k)$ (while $c_{66}$ serves as a cutoff) leading to a
$\ln z_J$,
\begin{eqnarray}\label{z3}
z_J\sim y_J\exp \{\frac{T}{8\pi}\int \frac {dk}{2\pi}\frac{(2\pi
d/a_0^2)^2}{c_{44}^0(k)+[\Sigma_1+z_c+I(k)]/k_z^2}\ln
z_J\}\nonumber
\end{eqnarray}
hence
\begin{equation}
T_d = \frac{4 a_0^4}{d^2} (\int_k \frac{dk}{c_{44}^{eff}(k)})^{-1}
\end{equation}
where an effective elastic modulus is
$c_{44}^{eff}(k)=c_{44}^0(k)+[\Sigma_1+z_c+I(k)]/k_z^2$. The $z_c$
term is obvious here by an expansion of ${\cal H}_{com}$ Eq.
(\ref{com}). However the $\Sigma_1+I(k)$ term cannot be derived by
an expansion of ${\cal H}_{dis}$ and a full RSB treatment is
required. $\Sigma_1$ acts then as $z_c$ and leads to a divergence
of  $c_{44}^{eff}(k)$ as $k\rightarrow 0$ as already noted in Ref.
\onlinecite{giam}. Since large $k$ dominates the integral in Eq.
(\ref{z3}) we use $I(k)\approx \Sigma_1+z_c$ so that
\begin{equation}\label{Td}
T_d\approx T_d^0[1+(\Sigma_1+z_c)\frac{8d\lambda_{ab}^2a_0^2}{\tau
\ln (a_0/d)}]
\end{equation}
where $T_d^0=\tau a_0^2\ln (a_0/d)/(4\pi \lambda_{ab}^2)$ from Eq.
(\ref{Td0}) is the transition temperature in the pure system.
Since $Q\sim 1/a_0$ and $W\sim a_0^{-4}$ we have from Eq.
(\ref{Sigma1}) $\Sigma_1+z_c\sim a_0^{-6}$. Thus the change in
$T_d$ due to columnar defects is $\delta T_d/T_d^0 \sim
a_0^{-4}\sim B^2$, up to $\ln B$ terms. From Eqs.
(\ref{Sigma1},\ref{Td}) we obtain our first principal result,
\begin{equation}\label{dTd}
\frac{\delta T_d}{T_d^0}\approx \frac{2(4\pi
)^3E_c^2n_{CD}b_0^4\lambda_{ab}^2}{\tau^2 a_0^4\ln (a_0/d)}\approx
10^2(\frac{b_0}{a_0})^4n_{CD}\lambda_{ab}^2
\end{equation}
where $E_c\approx 0.2\tau$ \cite{nelson}. For strong disorder and
strong fields the CD can dominate and then $T_d\sim B$ increases
with $B$. This allows a reentrant behavior, i.e. for a fixed
temperature as $B$ is increased a decoupling occurs at $T_d\approx
T_d^0\sim 1/B$ and then a recoupling would occur at a higher
field, assuming this field is still below melting.

Next we address the commensurability term, which, unlike the
Josephson coupling, depends also on the longitudinal $u_{l}({\bf
q},k)$ component. We therefore add longitudinal energy terms:
first an elastic energy of the form Eq. (\ref{el}) with $c_{66}$,
$c_{44}^0$ replaced by $c_{11}$ and $c_{44}^l$, respectively
\cite{goldin}, and secondly the usual long wavelength disorder
coupled to ${\mbox{\boldmath $\nabla$}}\cdot {\bf u}$ \cite{giam}
which yields the form of the first term of Eq. (\ref{dis}) with
$s$ replaced by $s^l$. Since $\sigma_{ab}$ originates from the $W$
term in Eq. (\ref{dis}) $\Sigma_1$ and $I(k)$ are common to both
longitudinal and transverse parts while the location of the one
step solution changes by a factor $c_{11}/(c_{11}+c_{66})$. Since
$c_{66}/c_{11}=a_0^2/16\pi \lambda_{ab}^2\ll 1$ the effect on
$\Sigma_1$ is small, yet the structure factor for the longitudinal
modes (analog of Eq. (\ref{Gc})) is significantly modified by the
same $\Sigma_1$ and $I(k)$.

The equation for $z_c$ depends also on the $k=0$ component of
$G_{aa}(q,k)$ which involves the long wavelength disorder
parameters $s\,,s^l$. Using inversion methods for $G_{ab}$
\cite{giam,mezard} we obtain our second prinicipal result,
\begin{equation}\label{zc}
z_c\sim y_c (\frac{z_c}{\Sigma_1
+z_c})^{1/2}(z_c)^{sQ^2/(16\pi
c_{66}^2)+s^lQ^2/(16\pi c_{11}^2)}\,.
\end{equation}
Hence at some critical $s\,,s^l$ (where the powers of $z_c$ on
both sides of (\ref{zc}) equal) the commensurability potential is
renormalized to zero. We note that this renormalization is driven
by $k=0$ terms, i.e. the same derivation is valid for a 2D system
with point disorder \cite{golub}.

Long wavelength disorder can generate dislocations
\cite{carpentier} at $s>c_{66}^2a_0^2/16\pi$, i.e. below the
critical value for the vanishing of $z_c$. Furthermore, on very
long scales dislocations will be induced by short wavelength CD
disorder as the system is effectively two-dimensional
\cite{carpentier,LG}. We limit our discussion to a Bragg glass
domain which ignores these very long scale effects.

The decoupling description neglects point defects, i.e. the
nucleation of VI. The latter were studied in the absence of
Josephson coupling and were shown to be generated by point
disorder \cite{HL}, leading to logarithmically correlated disorder
for VI . Disordered CD, however, induce only a $k=0$ component of
disorder which has exponentially decreasing correlations $\sim
(q^2+\lambda_{ab}^{-2})^{-2}(q^2+\Sigma_1)^{-1}$, i.e. a straight
flux line has exponentially decaying interactions even though each
of its pancake components has a logarithmic interaction. This type
of disorder cannot overcome the logarithmic interaction of pancake
vortices; hence, although for any finite (Gaussian) CD disorder
the ground state contains a finite density of VI perfectly aligned
along $z$, the defect transition temperature at which VI
uncorrelated between layers unbind is not affected by the CD. The
true decoupling, which allows for both Josephson phase
fluctuations and for point defects, lies in between the above
decoupling and the defect transition. For not too small Josephson
coupling the transition is near the decoupling one
\cite{horovitz3,HL} and therefore the results above should apply.

Finally, we address the data \cite{zeldov} on the melting curve
showing a kink at fields $B_k\gg B_{\phi}$. Within the proposed
porous vortex model \cite{zeldov} we suggest that the "vortex
matrix", pinned by the random CD, forms a commensurate potential. 
The lowest harmonic of this potential which couples 
to the flux periodicity has wavevector ${\bf Q}$; [harmonics 
with ${\bf Q}'>{\bf Q}$ have a $(z_c/(z_c+\Sigma_1))^{Q'^2/2Q^2}$ 
factor in Eq. 
(\ref{zc}), forcing a $z_c=0$ solution]. Since $s$ is a
second order effect, we consider $s^l=W/d^2\sim B_{\phi}B^2$,
hence decommensuration occurs at $B\sim B_{\phi}$, the bare
proportionality constant being, however, too small to account for the data.
The parameters $s, s^l$ are relevant parameters within RG
\cite{carpentier} so that their renormalized values can be large. The 
main result is then that elasticity dominates at large $B$ while 
disorder dominates at low B, driving $z_c\rightarrow 0$, in 
qualitative agreement with the data. We propose then to
search for an additional phase transition line within the solid
phase, corresponding to decommensuration, which meets the melting
curve at $B_k$.

In conclusion we have shown that columnar defects enhance the
decoupling transition so that $\delta T_d/T_d^0\sim B^2$. In
contrast, the melting temperature involves the same ratio, however
within a logarithm [see Eq. (\ref{B+})]; hence at weak disorder
the enhancement is also $\sim B^2$ while at strong disorder only a
weak $\ln B$ effect. The $B^2$ enhancement at strong disorder can
therefore be useful in identifying a decoupling transition.
Furthermore, for strong CD disorder a possibility of a reentrant
transition has been found, i.e. with increasing field decoupling
is followed by recoupling. We have also studied effects of a
commensurate CD density and shown that its potential vanishes
above a critical value of long wavelength disorder. This
decommensuration transition may account for the unusual kink in
the melting curve data.

We thank  E. Zeldov for valuable discussions. This
research was supported by THE ISRAEL SCIENCE FOUNDATION founded by
the Israel Academy of Sciences and Humanities and by a
German-Israeli DIP project.

\end{multicols}
\end{document}